# Cutting Soft Matter: Scaling relations controlled by toughness, friction, and wear†

Bharath Antarvedi Goda,[a] Zhenwei Ma,[b*] Stefano Fregonese,[a] and Mattia Bacca[a*]

Cutting mechanics of soft solids is gaining rapid attention thanks to its promising benefits in material characterization and other applications. However, a full understanding of the physical phenomena is still missing, and several questions remain outstanding. *E.g.*: How can we directly and reliably measure toughness from cutting experiments? What is the role of blade sharpness? In this paper, we explore the simple problem of wire cutting, where blade sharpness is only defined by the wire radius. Through finite element analysis, we obtain a simple scaling relation between the wire radius and the steady-state cutting force per unit sample thickness. The cutting force is independent of the wire radius if the latter is below a transition length, while larger radii produce a linear force-radius correlation. The minimum cutting force, for small radii, is given by cleavage toughness, *i.e.*, the surface energy required to break covalent bonds in the crack plane. The force-radius slope is instead given by the wear shear strength in the material. Via cutting experiments on polyacrylamide gels, we find that the magnitude of shear strength is close to the work of fracture of the material, *i.e.*, the critical strain energy density required to break a pristine sample in uniaxial tension. The work of fracture characterizes the toughening contribution from the fracture process zone (FPZ), which adds to cleavage toughness. Our study provides two important messages, that answer the above questions: Toughness can be estimated from wire-cutting experiments from the intercept of the force-radius linear correlation, as previously explored. However, as we discovered, this only estimates cleavage toughness. Additionally, the force-radius slope is correlated with the work of fracture, giving an estimation of the dissipative contributions from the FPZ.

## 1 Introduction

We encounter soft solid cutting in our everyday life and it is of enormous cultural and engineering importance with applications ranging from food processing[1], manufacturing[2], and robotic surgery[3]. Despite wide-ranging interest in cutting, we understand very little about cutting mechanics in soft materials because of the complexity arising from a combination of cutting tool geometry, contact conditions, and material properties[4]. Some earlier[5] and more recent[6,7] investigations on soft material cutting sought to understand the relationship between the cutting tool radius and fracture energy by designing thin Y-shaped samples, with the aim to control friction in cutting tests. Notably, by changing the boundary conditions at the crack tip, i.e., by tuning the opening angle of the Y-shaped sample, they observed a transition from the *True Cutting* regime to the *Tear Dominant* one. In the former, indentation of the crack tip dominates the process, while in the latter the crack propagates at a distance away from the tip of the cutting tool. Such a complex model system, however, challenges a simple understanding of the main energetic contributions to cutting, mainly due to dissipative mechanisms like fracture, friction, wear, and adhesion. This is the aim of our study, carried on with the simple model system of wire cutting.

In our study, we investigate the role -in wire cutting- of material properties such as fracture toughness $\Gamma$, shear modulus $\mu$, critical strain energy density or 'work of fracture' $W_f$, and strain stiffening coefficient $\alpha$. The latter controls the increment of rigidity under continued stretch in a soft material and is typical of biological tissue[8–10] and some synthetic soft materials[11,12].

The fracture toughness $\Gamma$ of a material provides the measure of its ability to resist fracture, while the shear modulus $\mu$ characterizes the material's resistance to deformation (indentation) at small strains. As expected, brittle materials require a lower force to propagate a crack and are therefore easier to cut. However, cutting resistance also depends on the stiffness of the material and dissipative mechanisms like friction[13,14]. This motivates careful consideration of $\Gamma$ in relation to $\mu$, $W_f$, and the dissipative forces introduced by friction, wear, and adhesion.

This paper provides important scaling relations between the steady-state cutting force, per unit sample thickness $T$, $F_{ss}/T$, and the sharpness of the blade, here defined as the wire radius $R_w$. We observe two important regimes, namely, (i) high toughness, or friction-dominated regime, at a relatively small wire radius, and (ii) low toughness, or wear-dominated regime, at larger radii. In regime (i), we observe that $F_{ss}/T$ does not depend on $R_w$, while in regime (ii), there is a linear $R_w - F_{ss}/T$ relationship.

[a] *Mechanical Engineering Department, University of British Columbia, Vancouver, BC V6T1Z4, Canada;*
[b] *Department of Pathology and Laboratory Medicine, University of British Columbia, Vancouver, BC V6T 2B5, Canada*
* E-mail: mbacca@mech.ubc.ca; zhenwei.ma@ubc.ca
† Electronic Supplementary Information (ESI) available: Supplementary Documentation. See DOI: 10.1039/cXsm00000x/

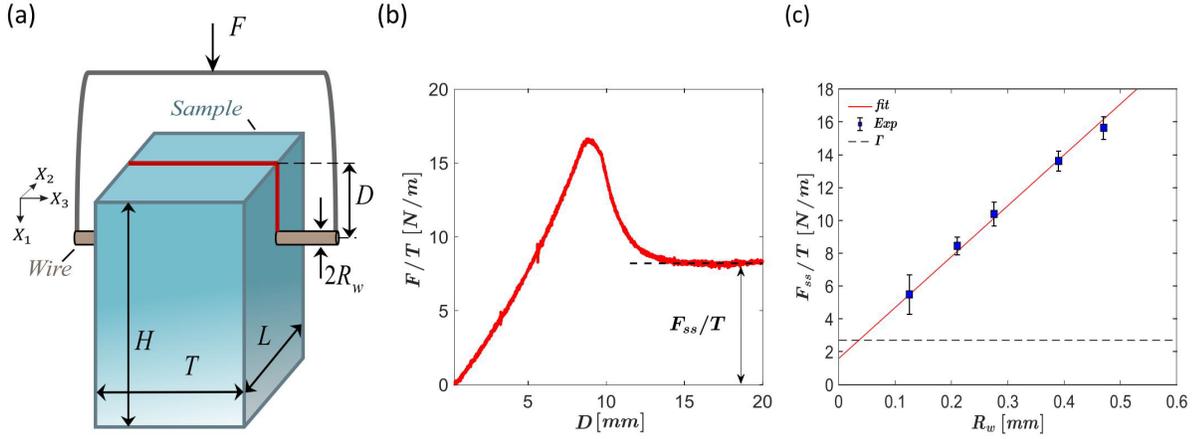

Fig. 1 (a) Schematics of the experimental setup adopted to perform wire cutting (b) Force per unit thickness $F/T$ versus wire depth $D$, highlighting the steady-state cutting force per unit thickness $F_{ss}/T$. (c) Relationship between $F_{ss}/T$ (mean of five cutting tests) vs wire radius $R_w$ for a PAAm hydrogel, comparing experiments (blue squares) with a linear fit (red solid line), and with indicated $\Gamma = 2.6 \; J/m^2$ (black dashed horizontal line).

The linear force-radius scaling we observe confirms previous observations on wire-cutting experiments by[4]. The minimum force, independent by radius, was instead observed by[6,7] on Y-shaped cutting experiments, and by[15] by cutting thin polymer films with biological blades. As explained in this paper, the transition wire radius $R_W^*$ between the two regimes, (i) and (ii), is proportional to the fracto-cohesive length of the material $l_f = \Gamma/W_f$[16]. In this manuscript, we first provide information about the wire-cutting experiments. Next, we describe our model system, and the hypothesis adopted to construct our finite element analysis (FEA). Finally, we provide our numerical results, fitted to obtain the force-radius scaling relations, ultimately confirming our experimental findings.

## 2 Experiments

We performed wire cutting experiments, as depicted in Fig. 1(a), using rigid steel wires with circular cross-sections to cut samples of polyacrylamide (PAAm) hydrogel. We purchased acrylamide (AAm, A8887), N,N'-Methylenebisacrylamide (MBAA, M7279), tetramethylethylenediamine (TEMED), and ammonium persulfate (APS) from Sigma Aldrich. The polyacrylamide (PAAm) hydrogel was prepared by mixing 10 mL of acrylamide solution (40% w/w), 1.5 mL of MBAA (2% w/w), 15 $\mu L$ of TEMED, and 1 mL of APS (66 mg/mL). The mixture was then injected into a closed acrylic mold overnight. Experimental characterization was done via uniaxial tensile tests and pure shear fracture tests. Uniaxial tension estimated the shear modulus $\mu \approx 685\,Pa$ and the work of fracture $W_f \approx 7\,kJ/m^3$ (the strain energy density required to break a pristine sample in uniaxial tension). Pure-shear fracture tests estimated toughness $\Gamma = 2.6\,N/m$. The SI provides more details related to the characterization of this material. We measured the force per unit thickness $F/T$ as a function of wire position $D$, as shown in Fig. 1(b). We focused on the steady-state cutting force $F_{ss}/T$ occurring after the initial force peak required to initiate the cut[17], as highlighted in Fig. 1(b). Fig. 1(c) reports the steady-state cutting forces $F_{ss}/T$ for each wire radius adopted, namely, $R_w = 0.125, 0.210, 0.275, 0.390$ and $0.470\,mm$. This figure reports each data point as the result from five trials, where the square indicates the median value and the error bar indicates the standard deviation. The tested material behaves like a neo-Hookean hyperelastic solid. The cutting speed is $\dot{D} = 0.2\,mm/s$. In Fig. 1(c), we can observe a linear correlation between $F_{ss}$ and $R_w$. The physical mechanism behind this phenomenon is explained in the next sections.

## 3 Model

We describe the hyperelastic behavior fo the cut sample with the incompressible Ogden strain energy density functional

$$W = \frac{2\mu}{\alpha^2}(\lambda_1^\alpha + \lambda_2^\alpha + \lambda_3^\alpha - 3) \quad (1)$$

where $\lambda_i$ are principal stretches and $\alpha$ characterizes strain stiffening tendency of the soft material[18]. $W$ depends on $\mu$, $\lambda$, and $\alpha$. Note that $\alpha = 2$ gives neo-Hookean hyperelasticity. The principal Cauchy stress in $X_1$ is

$$\sigma_1 = \lambda_1 \frac{\partial W}{\partial \lambda_1} - p \quad (2)$$

where p is the Lagrange multiplier enforcing incompressibility. In the example of wire cutting shown in Fig. 2(a), crack propagation is mediated by a cutting tool (the cylindrical wire), which loads the sample via contact stresses in the proximity of the crack (cut). This loading condition is substantially different from that of remote loading, like in the case of pure shear. Let us now analyze the *steady-state cutting force* $F_{ss}$ required to continuously cut the sample by pushing the wire through it. Consider the soft material specimen of length $L$ and thickness $T$ with a pre-existing cut of depth $c$, in which we insert the rigid wire of cross-sectional radius $R_w$ (Fig. 2(a)). Now, the work of cutting, per unit depth $D$ in $X_1$ direction, is

$$F_{ss}dD = \Gamma_o T dc + dU_s + dQ_d \quad (3)$$

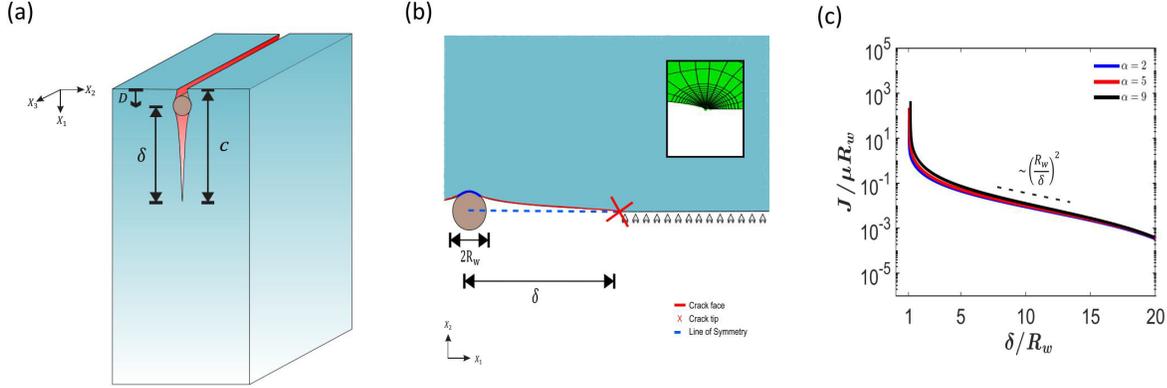

Fig. 2 (a) Schematics of wire cutting (along $X_1$-axis) of a soft material (in blue) with a pre-existing crack (in red) by a rigid wire (in brown) (b) FE half model (snippet showing mesh around the tip) of an Ogden hyperelastic material (in blue) with symmetry about X-axis was developed to evaluate the energy release rate, $J$ when the wire is inserted into a pre-existing crack. (c) Semi-log plot of dimensionless energy release rate $J/\mu R_w$ vs $\delta/R_w$ where $\Gamma_0$ is the toughness of the material $U_s$ is the stored elastic strain energy in the sample, and $Q_d$ is an additional mechanical energy dissipation, other than the energy required for crack propagation. We take the latter as the work done by dissipative forces like friction, adhesion, and damage.

Consider now $dD = 0$ so that the wire is not moving. In this case, we can neglect $dQ_d$ as all dissipation mechanisms activate only in the presence of wire motion. Thus, we have

$$\Gamma_o = -\frac{dU_s}{T dc} \quad (4)$$

*i.e.*, the elastic strain energy cumulated in the sample is immediately released via crack propagation, following Griffith's law. As shown in Fig. 2(b), we calculate the J-integral around the tip of the crack using finite element analysis (FEA) based on the configuration of the system, given by the ratio $\delta/R_w$, where $\delta = c - D$. Results from our FEA, as shown in Fig. 2(c), give

$$\frac{J}{\mu R_w} \sim \left(\frac{R_w}{\delta}\right)^2 \quad (5)$$

for $\delta > R_w$, i.e., $J < \mu R_w$, a regime that we call *far-field cutting*, while for $\delta \approx R_w$ we have $J \gg \mu R_w$, in the *indentation-cutting* regime, *i.e.*, when the wire touches the crack tip. In the far-field cutting regime, we observe a stress concentration that is akin to that from linear elastic fracture mechanics (LEFM), as shown in Fig. 3(a) from our FEA, with $\delta = 15 R_w$. In the indentation-cutting regime, on the other hand, we observe a stress concentration that is akin to that from nonlinear elastic fracture mechanics (NLEFM)[16], as shown in the same figure, from our FEA with $\delta = R_w$.

Now, because any increment in $U_s$ is immediately released to propagate the crack, for $D \gg R_w$, we can assume that $dc = dD$, i.e. $d\delta = 0$. In this scenario, the steady-state cutting force, from Eq. (3), becomes

$$F_{ss} = \Gamma_o T + F_d \quad (6)$$

where $F_d = dQ_d/dD$ is the resultant of all the dissipative forces projected along the wire path ($X_1$ axis). The latter depends on interfacial interactions between the wire and the specimen, like friction, adhesion, and the resistance of the material against wear. The force $F_d$ is then the integral of the projection of the (interfacial) contact shear stress $\tau_c$ along the $X_1$ direction, giving

$$F_d = 2 R_w T \int_{\theta_1}^{\theta_2} \tau_c \sin\theta \, d\theta \quad (7)$$

with

$$\tau_c = \xi P + \tau_w \quad (8)$$

as shown in Fig. 3(b). Here, $\xi P$ is the frictional shear stress transmitted at the interface, with $\xi$ the friction coefficient and $P$ the contact pressure. $\tau_w$ is the critical shear stress characterizing both the adhesive strength between the wire-specimen interface and the wear resistance of the material. Also, the angle $\theta$ (Fig. 3(b)) ranges between $\theta_1$ and $\theta_2$ since at larger or smaller angles the material is not in contact with the wire.

The magnitude of the dissipative force depends on the distribution of interfacial stresses between the wire and the specimen, which depends on the wire-specimen configuration given by the geometric ratio $\delta/R_w$. The latter, according to Fig. 2(c) and Eq. (5), is a function of the dimensionless parameter $\Gamma_o/\mu R_w$, where we have adopted the critical condition $J = \Gamma_o$.

## 4 Results

In this section, we analyze all contributions to the steady-state cutting force, assuming a very thick specimen, $T \gg R_w$, allowing for plane strain hypothesis. The correlations between the dimensionless dissipative force $F_d/\mu T R_w$, the dimensionless toughness $\Gamma_o/\mu R_w$, the friction coefficient $\xi$ and the dimensionless wear resistance $\tau_w/\mu$ are obtained from finite element analysis (FEA) and reported in Fig. 4. The numerical results in this figure can be fitted to the relation

$$\frac{F_d}{\mu R_w T} \approx \xi \left(\frac{\Gamma_o}{\mu R_w}\right)^n \quad (9)$$

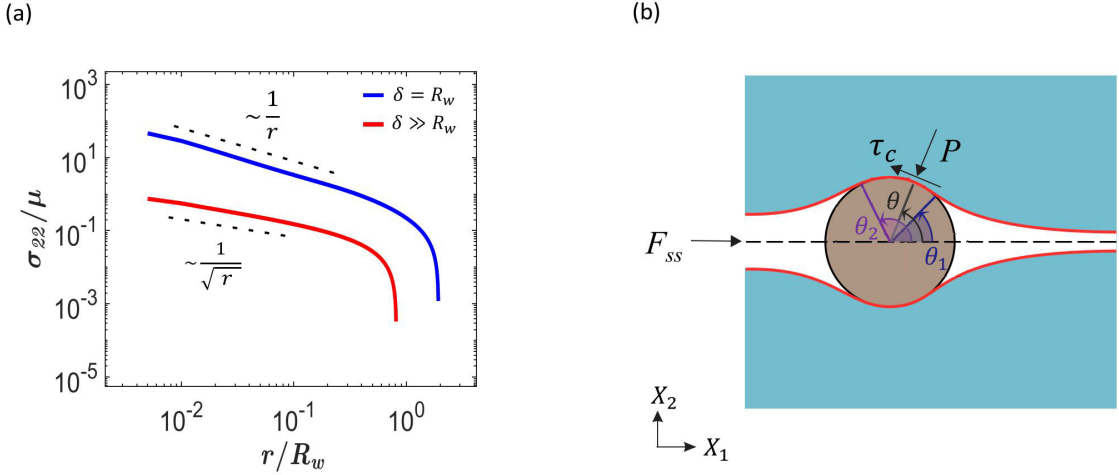

Fig. 3 (a) Log-log plot of dimensionless stress $\sigma_{22}/\mu$ vs dimensionless distance from the crack tip $r/R_w$ evaluated when $\delta = R_w$ (blue), giving nonlinear elastic fracture mechanics (NELFM) solution, and $\delta = 15 R_w$ ($\gg R_w$, red), giving linear elastic fracture mechanics (LEFM) solution (b) Schematics of wire cutting in the extreme regime of *high toughness*, where $\Gamma_o \gg \mu R_w$. Here, the power-law coefficient $n$ is reported in Table 1 as a function of the strain-stiffening coefficient $\alpha$. In this regime, $F_d$ appears to be independent of the wear resistance $\tau_w$, and mainly dependent on the strain-stiffening coefficient $\alpha$. Thus, the interfacial stresses in Eq. (8) are dominated by friction, given the high contact pressure $P$ due to the proximity of the wire with the crack tip ($\delta \approx R_w$). The high toughness as a friction-dominated regime was also observed in the puncture of soft solids[19]. From Table 1 we can deduce that $n \approx 1$, and thus that $F_d \approx \xi \Gamma_o T$, where the dissipative force is nearly independent of the wire radius $R_w$. Thus,

$$\frac{F_{ss}}{T}\bigg|_{\Gamma_o \gg \mu R_w} \approx (1+\xi)\Gamma_o \qquad (10)$$

In the other extreme regime of *low toughness*, $\Gamma_o \leq \mu R_w$, and the dissipative force $F_d$, from Fig. 4, can be fitted to the relation

$$\frac{F_d}{\mu R_w T} \approx 4\left(\xi + \frac{\tau_w}{\mu}\right) \qquad (11)$$

In this regime, as deduced from Eq. (11), we have a linear scaling between the dissipative force $F_d$ and the wire radius $R_w$. We also have negligible influence of strain stiffening, as shown in Fig. 4, where the interfacial forces in Eq. (8) are now dominated by the wear resistance of the specimen $\tau_w$, given the low contact pressure $P$ in this case. For this case, we have

$$\frac{F_{ss}}{T}\bigg|_{\Gamma_o \leq \mu R_w} \approx \Gamma_o + 4(\tau_w + \xi\mu)R_w \qquad (12)$$

describing the low toughness regime.

From Eq. (10) and (11), we can finally deduce the general cutting force equation

$$\frac{F_{ss}}{T} \approx (1+\xi)\Gamma_o + 4(\xi\mu + \tau_w)R_w \qquad (13)$$

representing all scenarios with good approximation, especially in the two cases analyzed above. In fact, Eq. (13) simplifies to (10) or (12), in the extreme regimes of high toughness (friction dominated), and low toughness (wear dominated), respectively.

Table 1 Power-law coefficient $n$ of the fitting function in Eq. (9), in the high toughness regime ($\Gamma_o \gg \mu R_w$) obtained to the least square feets with an r-square accuracy of $r^2 \geq 0.98$

| $\alpha$ | n |
|---|---|
| 2 | 1.18 |
| 5 | 1.06 |
| 9 | 0.96 |

By rearranging Eq. (13) we can obtain the dimensionless steady-state cutting force as

$$\frac{F_{ss}}{\Gamma_o T} \approx 1 + \xi + 4\left(\xi + \frac{\tau_w}{\mu}\right)\frac{R_w}{l_e} \qquad (14)$$

with $l_e = \Gamma_o/\mu$, the elasto-cohesive length of the material. Based on the above observations, the high-toughness regime applies when $R_w \ll l_e$, where the second term on the right-hand side of Eq. (14) becomes small, compared to the first one. Conversely, the low-toughness regime applies when $R_w \geq l_e$. However, it should be noted that, for most soft gels, $l_e$ is in the range of several *mm*, requiring relatively large radii to observe a linear scaling between cutting force and wire radius. As shown in Fig. 1(c), we observe such linear scaling for much smaller radii. A reasonable assumption is that the friction coefficient $\xi$ is negligibly small, thus $\xi \ll 1$. Under this assumption, Eq. (14) becomes

$$\frac{F_{ss}}{\Gamma_o T} \approx 1 + 4\frac{R_w}{l_w} \qquad (15)$$

with $l_w = \Gamma_o/\tau_w$ a length scale that correlates toughness with wear resistance. Suppose we assume $\tau_w \sim W_f$, given that wear is created by material damage, responsible for the development of a fracture process zone (FPZ). In that case, we have then $l_w \sim l_f$,

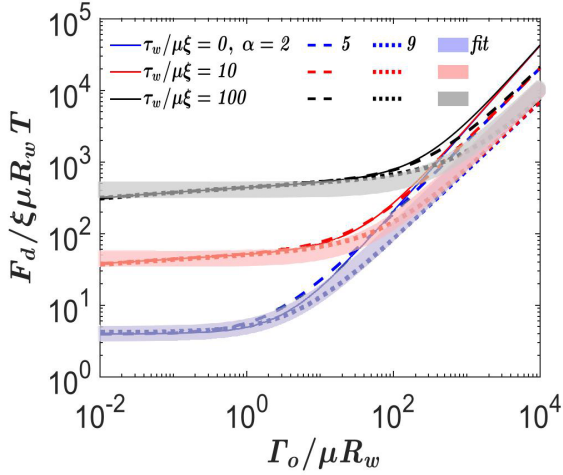

Fig. 4 Log-log plot of dimensionless dissipative force at steady-state $F_d/\xi\mu R_w T$ vs dimensionless energy release rate $\Gamma_o/\mu R_w$ at different values of $\tau_w/\mu\xi$ for $\alpha=2$, $\alpha=5$, and $\alpha=9$ using ABAQUS

with $l_f = \Gamma_o/W_f$, the fracto-cohesive length, which is typically in the range of 0.1 $mm$ for brittle gels (10 times smaller than $l_e$). Based on this assumption, from Eq. (15), we can conclude that the transition radius $R_w^* = l_f/4$ defines the force-radius scaling so that for $R_w \geq R_w^*$ we can observe the linear correlation.

In Fig. 1(c), we observe a linear correlation between $F_{ss}$ and $R_w$. Yet, the elasto-cohesive length of the material, based on the reported material properties in Section 2, is $l_e = 3.8\ mm$, thus, much larger than the wire radii. The fracto-cohesive length is $l_f = 371\mu m$, giving a transition radius of $R^* \approx l_f/4 = 92\mu m$, which is smaller than the smallest wire radius utilized in our experiments. In Fig. 1(c), the slope of the (red dashed) line connecting the experimental data points is 31 $kPa$, which, according to Eq. (10), neglecting friction, should be $4\tau_w$, giving $\tau_w = 7.75\ kPa$, surprisingly close to $W_f$. The intercept between the line connecting the data points and the y-axis gives an estimated $\Gamma_o = 1.6\ J/m^2$, which is smaller than the measured $\Gamma$ via pure-shear tests, as reported in the SI. Fig. 1(c) reports the horizontal (black dashed) line related to the measured $\Gamma$. We attribute the discrepancy between the toughness measured via pure shear tests and the intercept of the force-radius plot in Fig.1(c) to the additive contributions toward the measured toughness $\Gamma = \Gamma_o + \Gamma_d$, where $\Gamma_o$ is the cleavage toughness (adopted in our model), and $\Gamma_d$ is the additional contribution due to the formation of the FPZ[20]. Consistently, *Zhang et al.*[7] observed that the cutting force can be smaller than the tearing force or the toughness measured from pure-shear tests. Thus, our study provides more insight into the physical mechanisms differentiating cutting from tearing. While the former is driven by contact loading near the crack tip and dissipative mechanisms (friction and wear) are controlled by blade sharpness, the latter gives a macroscopic measure of toughness, and dissipative contributions (from the FPZ) that only depend on the material.

## 5 Discussion

The linear correlation between the cutting force $F_{ss}$ and the wire radius $R_w$, by neglecting friction, gives the simple $F_{ss}/T = \Gamma_o + 4\tau_w R_w$. By assuming that $F_{ss}/T \approx \Gamma$, where $\Gamma \approx \Gamma_0 + 4\tau_w R_w$, we can observe a similarity with the experiments performed by *Thomas et al.*[21], where uniaxial tensile tests of notched samples, created by razors of various tip radii $R$, gave a toughness scaling $\Gamma \sim W_f R$. In this analogy, the wear created by blunt razor blades has likely enhanced the formation of the FPZ, thus increasing the apparent toughness of the samples. However, more experimentation is needed to verify this hypothesis and establish the correlation between wear/damage (due to contact shear stresses) and the work of fracture determined from tensile tests for soft materials.

It should also be noted that wire cutting is highly affected by the three-dimensionality of the sample, especially at cut initiation[17]. However, in our steady-state cutting analysis, we assume the problem is plane strain. Owing to the low toughness of the sample, the three-dimensionality of the sample is likely to play a marginal role during steady-state cutting, and this is supported by the close validation between theory and experiments. For the reported experiments, we have a strong wear dominance, giving $\tau_w/\mu = 11.31 \gg \xi$, thus justifying the approximation in Eq. (15). This implies that the elasto-cohesive length must be larger than the fracto-cohesive one, specifically, $l_e \gg \xi l_w$. Given that in the tested materials $l_e/l_w \approx 10$, this means $\xi \ll 10$, satisfies the hypothesis if we assume $\xi < 1$. However, another condition is that of $\tau_w \ll \xi\mu$, for which, Eq. (14) would still provide a linear scaling between radius and force, but due to friction and stiffness instead of wear, giving $F_{ss}/T \approx \Gamma_o + 4\xi\mu R_w$. This would require $l_w/l_e \gg 1/\xi$. Thus, by assuming $\tau_w$ is due to wear/damage, for which $l_w \sim l_f$, the material, in this case, should have a much larger fracto-cohesive length than the elasto-cohesive one. As observed by *Long et al.*,[16] this is only the case in very stiff materials and not in soft elastomers and gels.

## 6 Conclusion

Our study explores the cutting mechanics of soft solids via the simple and minimalist problem of wire cutting, where blade sharpness is only defined by the wire radius $R_w$. We obtain a surprisingly simple scaling relation between the steady-state cutting force, per unit sample thickness $T$, $F_{ss}/T$, after cut initiation, and wire radius $R_w$. We observe two regimes, namely small wire radii at which we have scale indifference, and large radii, at which the cutting force scales linearly with wire radius. The former is defined by $R_w \ll R_w^*$, with $R_w^*$ a transition radius, while the latter applies when $R_w \geq R_w^*$. In the small radius regime, the cutting force only depends on toughness $\Gamma_o$ and the wire-sample friction coefficient $\xi$, where $F_{ss}/T = \Gamma_o(1+\xi)$. We call this a friction-dominated regime, and typically occurs at relatively high toughness. Also, in this regime, the wire is close to the crack tip and thus indents the sample while cutting it (indentation-cutting regime). Conversely, in the large radius regime, the cutting force depends on the toughness and the wear resistance (shear strength) of the material $\tau_w$ via $F_{ss}/T \approx \Gamma_o + 4\tau_w R_w$. This regime is defined as wear-dominated and typically occurs at a relatively low toughness. In this regime, the wire is far from the crack tip, giving 'far-field cutting'. From experimental validation of our results, we find $R_W^* = l_w/4$, where $l_w = \Gamma_o/\tau_w$ defining the interplay between threshold toughness and wear resistance. In our experiments, we observed that $\tau_w \approx W_f$, with $W_f$ the critical strain energy density at failure in uniaxial tensile tests of pristine samples. This suggests that $l_w \approx l_f$, so that $R_w^* \approx l_f/4$, and thus the transition radius is 1/4 of the fracto-cohesive length scale $l_f = \Gamma/W_f$. Our study provides two important messages: Toughness can be estimated from wire-cutting experiments from the intercept of the force-radius linear correlation, as previously found, but -as we discovered in this study- this method only provides the cleavage toughness $\Gamma_o$, i.e. the work of fracture at the net of dissipative contributions from the fracture process zone (FPZ); The linear slope of the $F_{ss}/T$ versus $R_w$ correlation is $4\tau_w \sim 4W_f$. In conclusion, wire-cutting experiments performed with various radii can provide an estimation of both cleavage toughness $\Gamma_o$ and the critical strain energy density $W_f$ controlling the dissipative contributions from the FPZ.

## Conflicts of interest

The authors declare no competing interests

## Acknowledgements

The work was supported by the Human Frontiers Science Program (RGY0073/2020), the Department of National Defense (DND) of Canada (CFPMN1–026), and the Natural Sciences and Engineering Research Council of Canada (NSERC) (RGPIN-2017-04464).

## Notes and references

# Supplementary Information: Cutting Soft Matter: Scaling relations controlled by toughness, friction, and wear[†]

Bharath Antarvedi Goda,[a] Zhenwei Ma,[b] Stefano Fregonese,[a] and Mattia Bacca[a*]

## 1 Experimental Measurements

### 1.1 Wire cutting experiments

Sample preparation, for cutting experiments, involves gluing the bottom of the hydrogels on a flat substrate fixed to the bottom clamp of the tensile testing machine (Instron, model 68SC-1), while the blade is a metal wire fixed onto the top clamp. The wire indents the sample moving orthogonally to its top surface, where the cut is initiated and then propagated. The wire moves at a constant prescribed velocity, while the force is recorded with a load cell (500 N).

### 1.2 Tensile tests of pristine samples: Evaluation of $\mu$ and $W_f$

Tensile rupture tests were performed on pristine "dog-bone" samples shown in Fig. S1. In the same figure, we also report the plot of the engineering stress (force over initial area) versus stretch (length over initial length) during the test. The plot reports a critical stress of $\sigma_c = 3.1\ kPa$. The area underneath the curve gives $W_f = 6.973 \pm 0.5\ kJ/m^3$ the critical strain energy density at failure. The stress-stretch plot was fitted with a neo-Hookean hyperelastic model prediction with an accuracy of $R^2 = 0.989$, unraveling a shear modulus of $\mu = 685 \pm 30\ Pa$.

Tensile test hydrogel samples are cut into dog-bone-shaped specimens (gauge length 30 mm, width 5 mm, and thickness 0.5 mm) and glued (Krazy Glue) to PET clamps. Samples are subject to uniaxial stretch by a tensile testing machine (Instron model 68SC-1) at a constant prescribed velocity (0.2 mm/s) while the force is recorded with a load cell (500 N). The engineering stress is obtained by dividing the force by the initial area of the sample cross-section.

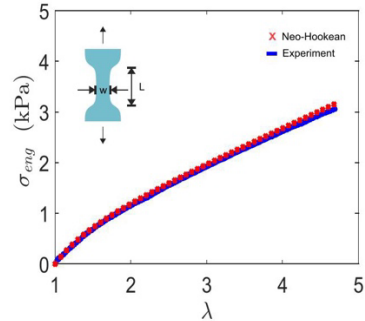

Fig. S1 Plot of engineering stress (force over the initial area) Vs stretch (length over initial length) for the uniaxial tensile test of a pristine sample of polyacrylamide (PAAm) hydrogel.

### 1.3 Pure shear fracture tests: Evaluation of $\Gamma$

Pure shear testing[1] is done using two samples of identical dimensions L = 45 mm, T = 1.4 mm, H = 5 mm, where one is pristine, and the other one has a notch of length c = 17.5 mm along the length. The pristine sample is pulled at a large stretch, while the notched one is pulled to its critical failure stretch prompting fracture. The critical strain energy density of the pristine sample at the critical fracture stretch of the notched one is $W_{ps}$ and is given by the area underneath the engineering stress versus stretch. We then have $\Gamma = W_{ps} H$, with $H$ the undeformed height of the sample. The estimated value for fracture toughness is $\Gamma = 2.6 \pm 0.2\ J/m^2$. Fig. S2 reports the engineering stress versus stretch for both the pristine sample (red line) and the notched one (blue line).

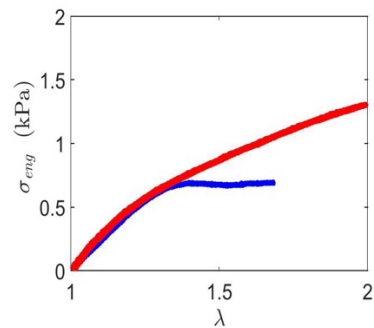

Fig. S2 Plot of engineering stress vs stretch for pure shear samples. Unotched sample (in red) and Notched sample (in blue)

[a] Mechanical Engineering Department, University of British Columbia, Vancouver, BC V6T1Z4, Canada;
[b] Department of Pathology and Laboratory Medicine, University of British Columbia, Vancouver, BC V6T 2B5, Canada
* E-mail: mbacca@mech.ubc.ca

## 2 Finite Element Analysis

The finite element analysis (FEA) was used to evaluate the J-integral and the cutting force contribution $F_D$ from wear and friction. From Eq. (7) we have $F_d/T = 2R_w \int_o^\pi \tau_c \sin\theta d\theta$, which, from Eq. (8) can be decomposed into

$$\frac{F_d}{2TR_w} = \tau_w \int_{\theta_1}^{\theta_2} \sin\theta \, d\theta + \xi \int_{\theta_1}^{\theta_2} P \sin\theta \, d\theta \quad \text{(S1)}$$

where $\theta_1$ and $\theta_2$ are the angles at which we have zero contact pressure ($P(\theta_1) = P(\theta_2) = 0$). By rearranging Eq. (S1) we have

$$\frac{F_d}{2TR_w} = \tau_w (cos\theta_1 - cos\theta_2) + \xi \frac{F_v}{TR_w} \quad \text{(S2)}$$

where $F_v = T \int_{\theta_1}^{\theta_2} P \sin\theta R_w d\theta$ is the vertical force applied to the wire to push it up into the half-specimen, as shown in Fig. 2(b).

Using FEA, we calculate $\theta_1$, $\theta_2$, and $F_v$ as functions of the configuration of the wire-specimen system, which is given by the ratio $R_w/\delta$. The simulation was performed on ABAQUS using R2D2 (a 2-node 2-D linear rigid link) for the cutting wire and CPE4H (a 4-node bilinear plane strain quadrilateral, hybrid, constant pressure) element to handle material incompressibility of the sample. Crack was assigned on the symmetry plane as an Engineering Feature with a crack extension direction (q-vector) to estimate the energy release rate.

### Notes and references

1 R. Rivlin and A. G. Thomas, *Journal of polymer science*, 1953, **10**, 291–318.